\title{Optimal Execution in a Multiplayer Model\\of Transient Price Impact}
\author{Elias Strehle\thanks{
                  Department of Mathematics, University of Mannheim.
                  The author gratefully acknowledges financial support by Deutsche Forschungsgemeinschaft (DFG) through Research Grant SCHI 500/3-2
                  and wishes to thank anonymous referees for helpful remarks.}
        }

\documentclass[11pt, a4paper]{article}
\usepackage[a4paper, left=2cm, right=2cm, top=2cm,bottom=2cm]{geometry}

\usepackage[round]{natbib}
\usepackage{tikz}
\usepackage{amsmath, amsfonts, amssymb, amsthm, mathrsfs, commath, mathtools}
\usepackage{enumitem}
\usepackage[T1]{fontenc}
\usepackage[latin1]{inputenc}
\usepackage{setspace}
\usepackage{commath}
\usepackage{mathtools}
\usepackage{bm}
\usepackage{bbm}
\usepackage{subfig}
\usepackage{color}
\usepackage{url}

\newtheorem{theorem}{Theorem}
\newtheorem{lemma}[theorem]{Lemma}

\newtheorem{remark}[theorem]{Remark}
\newenvironment{prf}[1][Proof.]{\begin{trivlist}
\item[\hskip \labelsep {\bfseries #1}]}{\end{trivlist}}

\DeclareMathOperator{\diag}{diag}

\newcommand{\Id}{I}
\newcommand{\overX}{\overline{X}}
\newcommand{\zero}{{\bm 0}}
\newcommand{\one}{{\bm 1}}

\renewcommand{\epsilon}{\varepsilon}
\renewcommand{\phi}{\varphi}

\begin{document}
\newpage
\maketitle

\begin{abstract}
	Trading algorithms that execute large orders are susceptible to exploitation by order anticipation strategies.
This paper studies the influence of order anticipation strategies in a multi-investor model of optimal execution under transient price impact.
Existence and unique\-ness of a Nash equilibrium is established under the assumption that trading incurs quadratic transaction costs.
A closed-form re\-pre\-sen\-ta\-tion of the Nash equilibrium is derived for exponential decay kernels.
With this representation, it is shown that while order anticipation strategies raise the execution costs of a large order significantly, 
they typically do not cause price overshooting in the sense of Brunnermeier and Pedersen.
\end{abstract}

The microstructure of financial markets has received much attention.
Advances in economic and mathematical theory have made it possible to drop the assumption of ``perfect'' financial markets, 
where trading is frictionless and liquidity infinite.
At the same time, the rise of high-speed algorithmic trading makes it increasingly important 
for investors and regulators to understand market microstructure
and be aware of the influence that transaction costs and liquidity have on the profitability of trading strategies and on the stability of financial markets.

Liquidity is particularly relevant for investors who must buy or sell large amounts of a financial asset over a relatively short period of time.
Trading large amounts in one piece can be prohibitively expensive or even impossible.
For large US stocks, the total volume of orders available in the limit order book at any given time 
is only about 1\% of daily trade volume \citep[p.~19]{Bouchaud2009}.
Hence a large order must usually be split into so-called child orders that are executed over the given time period.

In practice, the placement of child orders is typically handled by execution algorithms.
These al\-go\-rithms, including the popular VWAP (volume weighted average price, see for instance \citealp{Cartea2015}), 
are often based on the observation
that price impact depends on the relative size of an order: 
Price impact is smaller when markets are busy.
A simple execution algorithm might exploit this fact by trading every thirty seconds over the course of one trading day,
placing large positions when market volume is high and small positions when market volume is low;
while ensuring that the liquidation constraint is satisfied by the end of the day.
The major weakness of such a simple algorithm is its predictability.
Opportunistic investors with access to high-resolution data of financial markets (e.g., the entire limit order book)
can detect the algorithm and reverse-engineer it to predict its future trading behavior.
Then they can pursue an order anticipation strategy:
Trade in the same direction as the algorithm, but a little earlier;
then wait until the execution algorithm has traded and clear inventory directly afterwards.
This strategy---also known as front-running---turns the execution algorithm's price impact into a predictable source of profit.

Order anticipation strategies require sophisticated detection algorithms \citep{Hirschey2016} 
and a quick alternation of buy and sell orders.
Thus they are typically associated with high frequency traders, for instance by the \cite{SEC2010}.
Notice however that order an\-ti\-ci\-pa\-tion strategies do not require the breathtaking speed which is necessary for ``true'' high frequency strategies
such as stale order sniping or non-designated market making \citep{MacKenzie2011}.

That order anticipation strategies have been described as aggressive \citep{Benos2012}, predatory \citep{Brunnermeier2005} and ``algo-sniffing'' \citep{MacKenzie2011} suggests that the \cite{SEC2010} is not alone in suspecting that they ``may present serious problems in today's market structure'' (p.~3609).
Indeed, \cite{Tong2015} reports that ``one standard deviation increase in the intensity of [high frequency trading] activities 
increases institutional execution shortfall costs by a third''~(p.~4).

\cite{Brunnermeier2005} even suggest a direct connection between order anticipation strategies and financial breakdowns.
They argue that front-running amplifies the price drop caused by a large sell order, an effect known as price overshooting.
This might trigger further sell orders (e.g., from pending stop-loss orders), which are again subject to front-running,
causing further price overshooting and, ultimately, a complete market crash.

Even with high-quality data, empirical studies cannot perfectly identify order anticipation strategies in the market.
This is why it can be helpful to study them in a theoretical model.
This paper analyzes order anticipation strategies in a multi-investor model of optimal execution under transient price impact.
Special consideration is given to \citeauthor{Brunnermeier2005}'s (\citeyear{Brunnermeier2005}) 
claim that order anticipation strategies cause price overshooting.

The single-investor case of optimal execution under transient price impact was studied by \cite{Bouchaud2009} and \cite{Obizhaeva2013}.
Both assume that price impact decays exponentially, an assumption also made by \cite{Lorenz2013} and \cite{Schied2015a}.
Larger classes of decay kernels and their compatibility with absence of price manipulation are studied by 
\cite{Gatheral2010}, \cite{Gatheral2012} and \cite{Curato2017}.
\cite{Jaisson2015} and \cite{Alfonsi2016} link price impact with order flow modeled by a Hawkes process;
in this view, changes in the market price result from changed expectations about order flow imbalance.

If more than one investor trades, the problem of optimal execution can be analyzed with tools from game theory.
Multi-investor models of optimal execution include \cite{Carlin2007}, \cite{Schoneborn2009}, \cite{Carmona2011}, \cite{Moallemi2012}, \cite{Schied2015}, \cite{Lachapelle2016}, \cite{Cardaliaguet2016} and \cite{Huang2017}.

The model in this paper belongs to both areas of research: It features transient price impact with general decay kernels
and an arbitrary number of investors. 
Optimal execution strategies are characterized by Fredholm integral equations of the second kind,
similar to \cite{Gatheral2012}.
This observation also has wide-ranging implications in the single-investor case \citep{Schied2017}.

The paper is organized as follows.
Section~\ref{oas_chapter:existence} presents the model
and shows existence and uniqueness of a Nash equilibrium.
Each investor obtains his equilibrium strategy by solving a Fredholm integral equation of the second kind.

Section~\ref{oas_section:exponential} derives a closed-form representation 
of equilibrium strategies under the assumption that transient price impact decays exponentially.

Section~\ref{oas_section:orderanticipation} provides an economic analysis of order anticipation strategies 
based on the closed-form representation obtained in the previous section.
One investor liquidates a sell order, and~$n$ opportunistic investors pursue order anticipation strategies 
to benefit from the liquidating investor's price impact.
The influence of opportunistic investors on the liquidating investor's optimal strategy and expected costs is studied.
Furthermore, the claim by \cite{Brunnermeier2005} that opportunistic traders cause price overshooting is tested.
For many choices of parameters, it must be refuted.
In fact, opportunistic investors often produce the opposite effect and reduce the price drop caused by a sell order.
There are two possible explanations: 
Price overshooting does not occur if price impact is transient and sufficiently short-lived;
or price overshooting is prevented by quadratic transaction costs.

Section~\ref{oas_section:extensions} proposes an extension of the model in which opportunistic investors have additional time
to build up and unwind positions before and after the liquidating investor trades.

All proofs are in Appendix A.

\section{Existence of a Nash equilibrium}\label{oas_chapter:existence}
Consider a continuous time market for a single financial asset.
The asset is traded by~$n+1$ strategic investors over a time period~$[0,T]$.
In the absence of strategic trading, the asset price~$S^0$ is modeled as a right-continuous martingale 
on a filtered probability space~$(\Omega, \mathcal{F}, (\mathcal{F}_t)_{t \in [0,T]}, \mathbb{P})$ satisfying the usual conditions.
Assume that~$\mathcal{F}_0$ is~$\mathbb{P}$-trivial.

The strategic investors~$i=0, 1, \dotsc, n$ control their instantaneous rate of trading~$\alpha_i(t) \dif t.$
A positive sign of~$\alpha_i(t)$ corresponds to a buy order.
Each investor~$i$ must trade a fixed net amount~$x_i$ until time~$T$.
Consequently, a square-integrable function~$\alpha_i$ is called an \textit{(admissible) strategy} (for investor~$i$)
if it is progressively measurable and satisfies the \textit{liquidation constraint}~$\int_0^T \alpha_i(t) \dif t = x_i.$

Define the \textit{remaining net amount}~$X_i(t) \coloneqq x_i - \int_0^t \alpha_i(s) \dif s$.
In terms of~$X_i,$ the liquidation constraint reads~$X_i(T) = 0$.
Notice that~$\alpha_i$ and~$x_i$ together determine~$X_i$ and vice versa.
Therefore, an absolutely continuous function~$X_i \colon [0,T] \to \mathbb{R}$ will also be called an admissible strategy 
if~$X_i(0) = x_i$ and~$\alpha_i \coloneqq -\frac{\dif}{\dif t} X_i$ is an admissible strategy.

Every strategic investor impacts the asset price.
Price impact is assumed to be linear and transient and is modeled via a square-integrable \textit{decay kernel}~$G \colon [0, \infty) \to [0, \infty)$.
Suppose the strategic investors pursue strategies~$\alpha \coloneqq (\alpha_0, \alpha_1, \dotsc, \alpha_n).$
Similar to \cite{Gatheral2010}, the asset price evolves according to
\begin{equation}
\label{oas_eq:S}
	S(t) = S(t; \alpha) \coloneqq S^0(t) + \int_0^t G(t-s) \sum_{i=0}^n \alpha_i(s) \dif s.
\end{equation}

Investor~$i$'s costs from price impact are~$\int_0^T \alpha_i(t) S(t;\alpha) \dif t.$
In addition, each investor~$i$ incurs quadratic \textit{transaction costs}~$\frac{\gamma_i}{2} \alpha_i(t)^2 \dif t$, 
where~$\gamma_i \ge 0.$
Notice that the model explicitly allows for different levels of transaction costs for different investors.
In single-investor models, transaction costs of this form may be interpreted as costs arising from temporary price impact
\citep{Almgren2003,Bouchaud2009}.
It is tempting to follow \cite{Huang2017} in using the same interpretation for models with two or more investors.
But this is incorrect:
If an order generates temporary price impact at time~$t$,
it affects the execution price of every order subsequently executed at the same time~$t$.
It therefore becomes necessary to model the chronological order in which trades arriving at the same time are executed.
One might also choose to apply the same costs from temporary price impact to all orders arriving at the same time, as in \cite{Carlin2007}.
But notice that the probability of being executed first (and thus the probability of being subject to temporary price impact from other investors)
generally depends on the number of investors.
This must be taken into account when comparing models with different numbers of investors.
In any case, the transaction costs~$\frac{\gamma_i}{2} \alpha_i(t)^2 \dif t$ only affect the investor who caused them
and cannot be viewed as costs from temporary price impact.	
They should be interpreted as general costs arising from market frictions \citep[p.~751]{Gatheral2010} or a transaction tax \citep{Schied2015a}.
See \cite{Kissell2004} for a comprehensive overview of transaction costs on financial markets.

In total, investor~$i$ has the following \textit{costs of execution}:
\begin{equation}
\label{oas_eq:costs}
	J_i [ \alpha_i \,|\, \alpha_{-i} ] \coloneqq \int_0^T \big( \frac{\gamma_i}{2} \alpha_i(t)^2+ \alpha_i(t) S(t; \alpha) \big) \dif t,
\end{equation}
where~$
	\alpha_{-i} \coloneqq (\alpha_0, \dotsc, \alpha_{i-1}, \alpha_{i+1}, \dotsc, \alpha_n).
$

Assume that each investor is risk-neutral and therefore minimizes expected costs of execution.
Integration by parts shows that, for a given right-continuous martingale~$S^0,$
the term
$$
	\mathbb{E}\Big[ \int_0^T \alpha_i(t) S^0(t) \dif t \Big] =- x_i S^0(0) - \mathbb{E} \Big[ \int_0^T X_i(t) \dif S^0(t) \Big] =  - x_i S^0(0)
$$
is the same for all admissible strategies~$\alpha_i$.
Hence there is no loss of generality in assuming that~$S^0(t) = 0$ for all~$t \in [0,T]$.
Notice that this would be very different if traders were risk-averse \citep{Almgren2001}.

Assume further that all model parameters, including~$n$ and~$x \coloneqq (x_0, x_1, \dotsc, x_n)$, are known to each investor.
This is a standard---though often implicit---assumption in models with fixed liquidation constraints \citep{Carlin2007, Schoneborn2009, Carmona2011, Schied2015a}.
Two notable exceptions are the discrete time models in \cite{Moallemi2012} and \cite{Choi2015},
where investors try to derive each other's liquidation constraint from the evolution of the asset price.
This introduces a new strategic component: An investor might trade little in the beginning (or even make a feint and trade in the wrong direction)
to deceive others.
Modeling private information requires a sophisticated information structure: 
Each investor has his own filtration and updates his guesses throughout time in a Bayesian manner.
At the time, such an information structure seems unavailable for continuous time models.

The function~$\alpha^* = (\alpha^*_0, \alpha^*_1, \dotsc, \alpha^*_n)$ 
is called a \textit{Nash equilibrium (in the class of admissible strategies)} 
if for all~$i = 0, 1, \dotsc, n,$ the strategy~$\alpha^*_i$ is admissible, and~$
	\mathbb{E} [ J_i[\alpha_i^*; \alpha^*_{-i}] ] \le \mathbb{E} [ J_i[\alpha_i; \alpha^*_{-i}] ]
$
for every admissible strategy~$\alpha_i$ for investor~$i$.
In this case,~$\alpha^*_i$ is called an \textit{optimal strategy} (for investor~$i$).
Furthermore,~$\alpha^*$ is called a \textit{Nash equilibrium in the class of deterministic strategies} if each strategy~$\alpha^*_i$ 
is deterministic, and~$
	J_i[\alpha_i^*; \alpha^*_{-i}] \le J_i[\alpha_i; \alpha^*_{-i}]
$ 
for every deterministic admissible strategy~$\alpha_i$ for investor~$i$.

\begin{remark}
	The current analysis limits itself to (stochastic) \textit{open-loop strategies}~$\alpha_i(\omega,t)$, instead of \textit{closed-loop strategies} 
	$$
		\alpha_i \big(\omega, t, \alpha_0(\omega, t), \dotsc, \alpha_{i-1}(\omega, t), \alpha_{i+1}(\omega, t), \dotsc, \alpha_n(\omega, t)\big).
	$$
	In closed-loop Nash equilibria, investors still react optimally if another investor departs from equilibrium. 
	In open-loop Nash equilibria, this is typically not the case;
	each investor implicitly assumes that all other investors will pursue their respective equilibrium strategies.
	\cite{Carmona2011} show that this affects the equilibrium itself: 
	An open-loop Nash equilibrium need not be a closed-loop Nash equilibrium and vice versa.
	Closed-loop Nash equilibria are an appealing concept, but notoriously difficult to find. 
	See nonetheless the aforementioned paper for numerical simulations of open-loop and closed-loop 
	Nash equilibria in a model of optimal execution under temporary price impact.
	For a detailed discussion of open-loop and closed-loop equilibria 
	in the context of stochastic differential games, see Section~2.2 in \cite{Yeung2006}.
	
	Notice that the decision to search for open- or closed-loop equilibria is independent of the information structure of the model.
	In particular, a closed-loop equilibrium might still rely on the assumption that all liqudation constraints are known to each investor.
\end{remark}

\label{oas_page:positivetype}
Not every decay kernel~$G$ is sensible from an economic point of view.
Suppose there is only one investor~$i = 0$ and assume for simplicity that~$\gamma_0 = 0.$
The investor's costs from price impact are
\begin{equation}
\label{oas_eq:positivetype}
	\int_0^T \alpha_0(t) S(t; \alpha_0) \dif t 
		= \frac12 \int_0^T \int_0^T G(|t-s|) \alpha_0(t) \alpha_0(s) \dif s \dif t.
\end{equation}
If there is a strategy~$\alpha_0$ that makes~\eqref{oas_eq:positivetype} negative, the decay kernel~$G$ admits \textit{price manipulation} in the sense of \cite{Huberman2004}:
The investor can exploit his own price impact to generate arbitrarily large expected profits.
\cite{Gatheral2010} points out that price manipulation strategies do not constitute classical arbitrage 
because their profitability is affected by random fluctuations in the asset price.
They belong to the larger class of statistical arbitrage strategies which on average earn excess returns.
Notice the difference between price manipulation and order anticipation: 
A price manipulation strategy ge\-ne\-rat\-es profits from its own price impact,
an order anticipation strategy generates profits from another investor's price impact.
\cite{Alfonsi2016} explore the connection between absence of price manipulation and absence of opportunistic trading in more depth (Corollary~5.4).

If~\eqref{oas_eq:positivetype} is nonnegative for every~$\alpha_0 \in L^2[0,T]$ and every~$T > 0,$ 
then~$G$ is said to be of \textit{positive type} \citep{Mercer1909}.
Assume from now on that~$G$ is of positive type.

In models of optimal execution under transient price impact without transaction costs, 
impulse trades---i.e., jumps in~$X_i$---are optimal if transaction costs are zero \citep{Gatheral2012, Obizhaeva2013,Schied2015b}.
But in the current model, such jumps are inadmissible as~$X_i$ is required to be absolutely continuous.
This suggests that no Nash equilibrium exists as soon as~$\gamma_i = 0$ for some~$i$.
Assume from now on that~$\gamma_i > 0$ for all~$i = 0, 1, \dotsc, n.$
This assumption can be weakened in the single-investor case, see \cite{Schied2017}.

Under these assumptions on~$G$ and~$\gamma_i$, uniqueness of a Nash equilibrium 
is a simple consequence of the convexity of the cost functionals $J_i$.
The following results are easily adapted from \cite{Schied2015b}.
Uniqueness should be understood as uniqueness~$\mathcal{B}([0,T]) \otimes \mathbb{P}$-almost everywhere.

\begin{lemma}[Proposition~4.8 and Lemma~4.9 in \citealp{Schied2015b}]
\label{oas_lemma:zhang}
~
	\begin{enumerate}[label=(\roman*)]
		\item There is at most one Nash equilibrium in the class of admissible strategies.
		\item A Nash equilibrium in the class of deterministic strategies is also a Nash equilibrium in the class of admissible strategies.
	\end{enumerate}
\end{lemma}

The next step is to show that Nash equilibria are characterized by $n+1$ Fredholm integral equations of the second kind.
Existence of a Nash equilibrium then follows from the invertibility of the corresponding integral operator.

For~$\eta \in \mathbb{R}^{n+1},$ let~${\bm \eta}$ denote the~$(n+1)$-dimensional constant function~${\bm \eta}(t) = \eta.$
With slight abuse of notation, let~${\bm 0}$ and~${\bm 1}$ denote the~$(n+1)$-dimensional constant functions~${\bm 0}(t) = (0,0, \dotsc, 0)$ 
and~${\bm 1}(t) = (1,1, \dotsc, 1).$
Define a diagonal matrix~$\Gamma \coloneqq \diag(\gamma_0, \gamma_1, \dotsc, \gamma_n)$
and an operator~$F$ on~$L^2([0,T];\mathbb{R}^{n+1})$ via
\begin{equation}
\label{oas_eq:F}
\begin{aligned}
	(F \alpha)(t) &\coloneqq \Gamma \alpha(t) + \Big(\int_0^t G(t-s) \alpha(s)^\top \one(s) \dif s\Big)\, \one(t) \\
		&\hphantom{{}=}{}+ \int_t^T G(s-t) \alpha(s) \dif s.
\end{aligned}
\end{equation}

%

The following lemma connects the operator~$F$ with equilibrium strategies.

\begin{lemma}\label{oas_lemma:fredholm}
	The function~$\alpha^* = (\alpha^*_0, \alpha^*_1, \dotsc, \alpha^*_n) \in L^2([0,T];\mathbb{R}^{n+1})$ is a Nash equilibrium 
	in the class of deterministic strategies if and only if
	\begin{enumerate}[label=(\roman*)]
		\item $\int_0^T \alpha^*_i(t) \dif t = x_i$ for every~$i=0, 1, \dotsc, n,$ and
		\item there is an~$\eta = (\eta_0, \eta_1, \dotsc, \eta_n) \in \mathbb{R}^{n+1}$ such that~$(F \alpha^*)(t) = \eta$ 
			for almost all~$t \in [0,T]$.
	 \end{enumerate}
	In this case,~$\eta_i x_i \ge J_i[\alpha^*_i \,|\, \alpha^*_{-i}]$ for every~$i = 0, 1, \dotsc, n.$
\end{lemma}

\begin{remark}
	The optimality condition~$(F\alpha^*)(t) = \eta$ can be rewritten as
	\begin{equation}
	\label{oas_eq:fredholmremark}
		\gamma_i \alpha^*_i(t) + \int_0^T G(|t-s|) \alpha^*_i(s) \dif s = \eta_i - \int_0^t G(t-s) \sum_{j \neq i} \alpha_j^*(s) \dif s
	\end{equation}
	for~$i = 0, 1, \dotsc, n.$
	For fixed~$(\alpha^*_0, \dotsc, \alpha^*_{i-1}, \alpha^*_{i+1}, \dotsc, \alpha^*_{n})$, this
	is a one-di\-men\-sio\-nal \textit{Fredholm integral equation of the second kind}.
	Fredholm in\-te\-gral equations are connected with the more prevalent Euler-Lagrange equations in the following way:
	Consider the constant decay kernel~$G(t) = 1.$ 
	Then for admissible strategies~$\alpha = (\alpha_0, \alpha_1, \dotsc, \alpha_n)$
	the asset price evolves according to
	$$
		S(t) = \int_0^t \big( \alpha_i(s) + \sum_{j \neq i} \alpha_j(s) \big) \dif s.
	$$
	This turns the minimization of expected costs~$J_i[\alpha_i; \alpha_{-i}]$ into a classical problem in the calculus of variations.
	The corresponding Euler-Lagrange equation characterizing the optimal strategy~$\alpha^*_i$ is
	$$
		0 = \frac{\dif}{\dif t} \big[ \gamma_i \alpha^*_i + S \big] - \alpha^*_i.
	$$
	A straightforward calculation shows that this is the~$t$-derivative of~\eqref{oas_eq:fredholmremark}.
	But as soon as the decay kernel~$G$ is not constant, the nonlocal term~$\int_t^T G(s-t) \alpha^*_i(s) \dif s$ prevents the derivation
	of a ``proper'' (i.e., local) Euler-Lagrange equation from~\eqref{oas_eq:fredholmremark}.
	One might suspect that the minimization of expected costs can still be performed with classical methods
	by considering the two-dimensional process~$(\alpha^*_i,S)$ instead of~$\alpha^*_i$. 
	But~$S$ is a function of~$\alpha^*_i$ and a ``chain rule'' applies in the derivation of the Euler-Lagrange equation. 
	The additional term introduced by the chain rule 
	is just~$\int_t^T G(|t-s|) \alpha^*_i(s) \dif s$ \citep{Avron2003}.
	Hence Fredholm integral equations, not Euler-Lagrange equations, are the appropriate tool 
	for solving problems of optimal execution under transient price impact.
\end{remark}

Existence of a Nash equilibrium in the class of deterministic strategies is now shown by proving that~$F$ is invertible
and invoking the uniqueness result from Lemma~\ref{oas_lemma:zhang}.

\begin{theorem}\label{oas_th:existence}
	There is a unique Nash equilibrium~$\alpha^*$ in the class of admissible strategies.
	It is deterministic.
\end{theorem}

Recall that Theorem~\ref{oas_th:existence} relies on the following assumptions: The decay kernel~$G$ is square-integrable, $G(0)$ is bounded 
and each investor incurs strictly positive transaction costs $\gamma_i > 0.$
It seems likely that the first two assumptions could be relaxed without affecting the result.
In the single-investor case, \cite{Schied2017} establish existence of an optimal strategy for all convex, nonincreasing and integrable decay kernels 
(including those with a weak singularity $\lim_{t \to 0} G(t) = \infty$).

Assuming strictly positive transaction costs, however, is crucial.
The single-investor case without transaction costs is studied extensively by \cite{Gatheral2012}.
Here, optimal strategies need no longer be absolutely continuous (i.e., the remaining net amount~$X^*$ need not be differentiable),
which complicates the analysis considerably.
The authors show that no optimal strategy exists for the Gaussian decay kernel~$G(t) = e^{-t^2}$, although it is square-integrable and bounded (Example~2.16).
Consider also Theorem~4.5 in \cite{Schied2015b}, where the existence of a Nash equilibrium depends on the level of transaction costs.

\section{Transient price impact with exponential decay}
\label{oas_section:exponential}

\begin{figure}
	\centering
	\begin{tikzpicture}
		\node[anchor=south west,inner sep=0] (image) at (0,0) {\includegraphics[width=0.5\textwidth]{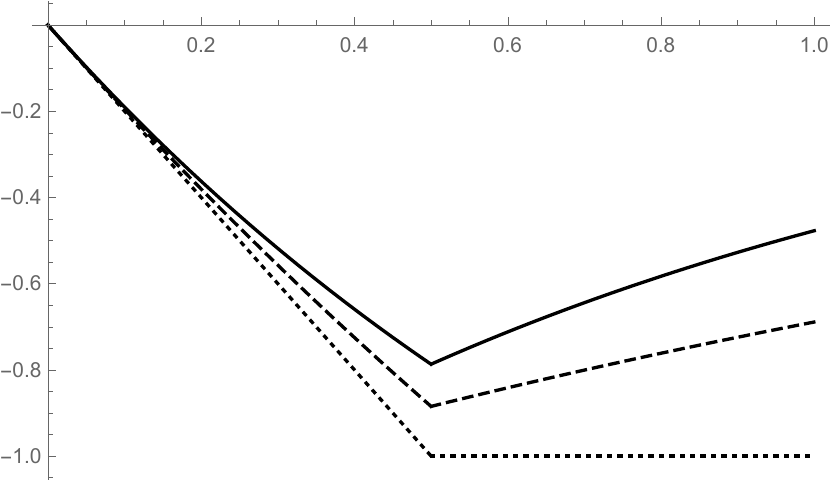}};
	    	\begin{scope}[x={(image.south east)},y={(image.north west)}]
			\node at (0.063,-0.07) {$S(t)$};
        		\node at (1.03,0.95) {$t$};
    		\end{scope}
	\end{tikzpicture}
	\caption{Illustration of transient price impact with exponential decay $G(t) = e^{-\rho t}$. 
		Asset price~$S(t)$ for~$\rho = 1$ (solid line),~$\rho = 0.5$ (dashed line) 
		and for permanent price impact~$\rho = 0$ (dotted line). 
		Here, a single strategic investor trades at a constant rate~$\alpha_0(t) = -2$ while~$t \le 1/2$ and~$\alpha_0(t) = 0$ while~$t > T/2$.
		Parameters:~$n = 0$, $T = 1$ and $x_0 = -1.$
		Notice that~$\alpha_0$ is admissible but not optimal.}
	\label{oas_fig:singleinvestor_expdecay}
\end{figure}
	
To obtain an explicit representation of the Nash equilibrium, assume from now on that transient price impact decays at an exponential rate, 
i.e.,~$G(t) = e^{-\rho t}$, as in \cite{Obizhaeva2013} and~\cite{Schied2015b}.
The parameter~$\rho > 0$ determines the size and persistence of price impact.
A small~$\rho$ implies large impact and slow recovery (see Figure~\ref{oas_fig:singleinvestor_expdecay}).
The limit~$\rho = 0$ corresponds to permanent price impact as in \cite{Almgren2001}.

In searching for a closed-form representation of the Nash equilibrium, it will turn out that the~$(n+2)$-dimensional function 
$$
	\psi \coloneqq (\alpha_0, \alpha_1, \dotsc, \alpha_n, S)
$$ 
is a more natural object of study than~$\alpha \coloneqq (\alpha_0, \alpha_1, \dotsc, \alpha_n).$
Clearly,~$\psi$ determines~$\alpha$ and vice versa.
With a slight abuse of terminology, the function~$\psi^* = (\alpha^*_0, \alpha^*_1, \dotsc, \alpha^*_n, S^*)$ 
will also be called a Nash equilibrium if~$\alpha^* \coloneqq (\alpha^*_0, \alpha^*_1, \dotsc, \alpha^*_n)$ 
is a Nash equilibrium and~$S^* = S(\cdot\,; \alpha^*)$.

Notice that~$G$ is of positive type.
Assume again~$\gamma_i > 0$ for all~$i = 0, 1, \dotsc, n.$
Let~$\alpha^* = (\alpha^*_0, \alpha^*_1, \dotsc, \alpha^*_n)$ be the unique Nash equilibrium
(compare Lemma~\ref{oas_lemma:zhang} and Theorem~\ref{oas_th:existence}).
It is deterministic and continuous.
Let~$\eta = (\eta_0, \eta_1, \dotsc, \eta_n)$ denote the corresponding vector for which~$F\alpha^* = {\bm \eta}$ 
(compare Lemma~\ref{oas_lemma:fredholm}).
Finally, let~$S^* \coloneqq S(\,\cdot\,; \alpha^*).$

The following approach yields a closed-form representation of $\psi^*$ and thus of $\alpha^*.$
Rewrite the optimality conditions in Lemma~\ref{oas_lemma:fredholm} as a system of integral equations:
\begin{equation}
\label{oas_eq:fredholmexp}
	\gamma_i \alpha_i^* (t) + S^*(t) + \int_t^T e^{\rho(t-s)} \alpha^*_i(s) \dif s = \eta_i,
	\quad t \in [0,T],
\end{equation}
for~$i = 0, 1, \dotsc, n.$

If all investors had homogeneous transaction costs~$\gamma_0 = \gamma_1 = \dotsb = \gamma_n$,
one could sum~\eqref{oas_eq:fredholmexp} over~$i$ to obtain a two-dimensional system of differential equations 
characterizing~$\sum_{i=0}^n \alpha^*_i$ and~$S^*$.
Once this system were solved,~\eqref{oas_eq:fredholmexp} would reduce to~$n+1$ identical one-dimensional ordinary differential equations.
The model in \cite{Schied2015} allows for this approach.
But if transaction costs are heterogeneous, all functions~$\alpha^*_0, \alpha^*_1, \dotsc, \alpha^*_n$ and~$S^*$ 
must be computed simultaneously.

Let~$i = 0, 1, \dotsc, n.$
It is clear from~\eqref{oas_eq:fredholmexp} that~$\alpha^*_i$ is differentiable in~$t$.
Differentiating and plugging in from~\eqref{oas_eq:fredholmexp} yields the ordinary differential equation
\begin{align}
\label{oas_eq:secondorderode}
	\frac{\dif}{\dif t} \alpha^*_i = \rho \alpha^*_i - \frac1{\gamma_i} \sum_{j\neq i} \alpha^*_j + \frac{2\rho}{\gamma_i} S^* - \frac{\rho \eta_i}{\gamma_i}.
\end{align}
Furthermore,~\eqref{oas_eq:S} shows that~$S^*$ satisfies the ordinary differential equation 
\begin{equation}
\label{oas_eq:Sderivative}
	\frac{\dif}{\dif t} S^* = \sum_{i=0}^n \alpha^*_i - \rho S^*.
\end{equation}

Combine~\eqref{oas_eq:secondorderode} and~\eqref{oas_eq:Sderivative} to conclude that the Nash equilibrium
$$
	\psi^* \coloneqq (\alpha^*_0, \alpha^*_1, \dotsc, \alpha^*_n, S^*)
$$ 
solves a system of differential equations of the form~$
	\frac{\dif}{\dif t} \psi^* = M \psi^* + m,
$
where~$M$ is a square matrix and~$m$ is a vector.

Let~$e^{Mt}$ denote the matrix exponential of~$Mt$.
If~$M$ is invertible, then~$\psi^*$ must be of the form~$\psi^*(t) = e^{Mt}z - M^{-1} m$ for some~$z \in \mathbb{R}^{n+2}.$
Notice that the investors' liquidation constraints translate into unusual boundary conditions for~$\psi^*$: 
They apply to~$\int_0^T \alpha_i^*(t)\dif t$, not~$\alpha^*_i$.
This complicates the calculation of~$z.$

The next theorem contains the closed-form representation of~$\psi^*$.
Define the~$(n+2)$-dimensional square matrices
\begin{align*}
	M &\coloneqq
	\begin{pmatrix}
		\rho & -\frac1{\gamma_0} & \cdots & -\frac1{\gamma_0} & \frac{2\rho}{\gamma_0} \\
		-\frac1{\gamma_1} & \rho & \cdots & -\frac1{\gamma_1} & \frac{2\rho}{\gamma_1} \\
		\vdots & \vdots & \ddots & \vdots & \vdots \\
		-\frac1{\gamma_n} & -\frac1{\gamma_n} & \cdots & \rho & \frac{2\rho}{\gamma_n} \\
		1 & 1 & \cdots & 1 & -\rho
	\end{pmatrix}
\intertext{and}
	N_1 &\coloneqq
	\begin{pmatrix}
		\rho \gamma_0 & 0 & \cdots & 0 & \rho \\
		0 & \rho \gamma_1 & \cdots & 0 & \rho \\
		\vdots & \vdots & \ddots & \vdots & \vdots \\
		0 & 0 & \cdots & \rho \gamma_n & \rho \\
		\gamma_0 & \gamma_1 & \cdots & \gamma_n & n+1
	\end{pmatrix}.
\end{align*}

Define further the~$(n+1)\times(n+2)$-dimensional matrix
$$
	W \coloneqq \begin{pmatrix}
		1 & 0 & \dots & 0 & 0 \\
		0 & 1 & \dots & 0 & 0 \\
		\vdots & \vdots & \ddots & \vdots & \vdots \\
		0 & 0 & \dots & 1 & 0
	\end{pmatrix}
$$
and the~$(n+2)$-dimensional column vector~$v \coloneqq (0, 0, \dotsc, 0, 1).$
For~$m \in \mathbb{N} \setminus \{0\},$ denote by~$\Id_m$ the~$m$-dimensional identity matrix.
Define the~$(n+2)$-dimensional square block matrix
$$
	N_2 \coloneqq \begin{bmatrix} W \big((M^{-1} + N_1 T) e^{MT} - M^{-1}\big) \\ v^\top  (\Id_{n+1} + N_1 e^{MT}) \end{bmatrix}.
$$

\begin{theorem}
\label{oas_th:ode}
	The matrix~$N_2$ is invertible and it holds that
	\begin{equation}
	\label{oas_eq:optimalstrategies}
		\psi^*(t) = (e^{Mt} + N_1 e^{MT}) N_2^{-1} \tilde{x}, 
	\end{equation}
	where~$\tilde{x} \coloneqq (x_0, x_1, \dotsc, x_n, 0)$.
\end{theorem}
	
The Nash equilibrium~\eqref{oas_eq:optimalstrategies} can be approximated numerically.
The next section does so and studies how opportunistic investors affect optimal strategies.

\begin{remark}
	The approach that leads to Equation~\eqref{oas_eq:optimalstrategies} does not generalize to other decay kernels.
	One of the first steps is to differentiate the optimality condition~$(F\alpha^*)(t) = \eta.$
	This yields
	$$
		0 = (F\alpha^*)'(t) = \gamma_i \frac{\dif}{\dif t} \alpha_i^*(t) + \int_0^t G'(t-s) \alpha_i^*(s) \dif s - \int_t^T G'(s-t) \alpha_i^*(s) \dif s.
	$$
	For exponential decay kernels, the relationship $G'(t) = -\rho\, G(t)$ leads to the ordinary differential equation~\eqref{oas_eq:secondorderode}.
	For all other decay kernels, it is unclear how to proceed.
	
	Even in the single-investor case $n=0$, closed-form solutions for non-exponential decay kernels are challenging to derive 
	\citep[Examples~6~and~7]{Schied2017}.
\end{remark}

\section{Economic analysis}
\label{oas_section:orderanticipation}

Let~$n \ge 1.$
Assume that investor~$0$ executes a net sell order~$x_0 < 0,$ while all other investors~$i = 1, 2, \dotsc, n$
trade zero net amounts~$x_i = 0.$
The case~$x_0 > 0$ is perfectly symmetric.
Investors~$i = 1, 2, \dotsc, n$ will only trade
if they can generate a profit (that is, negative costs) from the price impact generated by investor~$0$.
In this sense, they are \textit{opportunistic investors}.
Investor~$0$ will be referred to as the \textit{liquidating investor}.

Assume for simplicity that all opportunistic investors have identical levels of transaction costs, i.e.,~$\gamma_1 = \gamma_2 = \dotsb = \gamma_n.$
It follows from Lemma~\ref{oas_lemma:zhang} that in equilibrium, all opportunistic investors pursue the same strategy~$\alpha^*_1$.
Hence~$\overline{\alpha} \coloneqq \sum_{i=1}^n \alpha^*_i$ equals~$n \alpha^*_1.$
 	 
For the subsequent analysis, it is most illustrative to study the remaining net amounts
$$
	X^*_0(t) = x_0 - \int_0^t \alpha^*_0(s) \dif s
	\qquad\text{ and }\qquad
	\overline{X}(t) \coloneqq - n\int_0^t \alpha^*_1(s) \dif s.
$$

\begin{figure}
	\centering
	\begin{tikzpicture}
		\node[anchor=south west,inner sep=0] (image) at (0,0) {\includegraphics[width=0.5\textwidth]{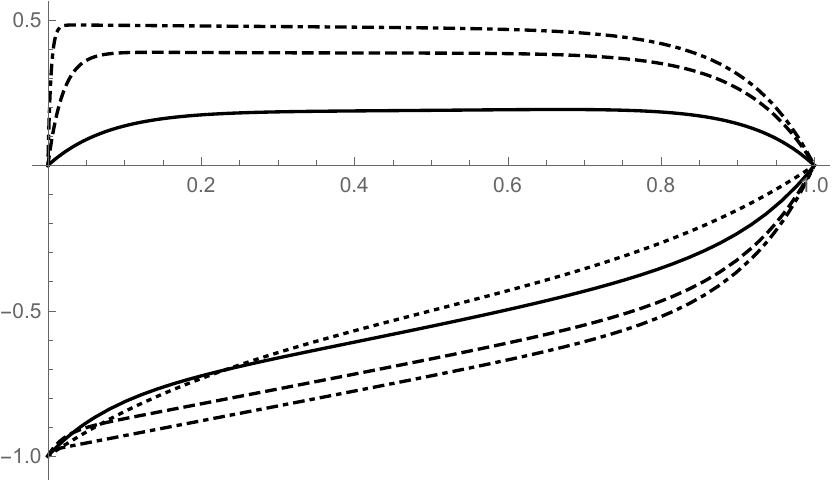}};
	    	\begin{scope}[x={(image.south east)},y={(image.north west)}]
			\node at (0.053,1.07) {$X_0^*(t), \overline{X}(t)$};
        		\node at (1.035,0.665) {$t$};
    		\end{scope}
	\end{tikzpicture}
	\caption{
		Remaining net amounts~$X^*_0(t)$ and~$\overX(t)$ 
		for~$n=0$ (dotted line),~$n=1$ (solid line),~$n=5$ (dashed line) and~$n=25$ (dot-dashed line).	
		Parameters:~$T = 1$, $\rho = 0.95$, $\gamma_0 = \gamma_1 = 0.1$, $x_0 = -1$ and $x_1 = 0.$
		}
	\label{oas_fig:npredator_highlowtransaction}
\end{figure}

Figure~\ref{oas_fig:npredator_highlowtransaction} is representative of equilibrium strategies in general.
Opportunistic investors engage in \textit{front-running}: 
They build up short positions in the beginning and buy back for the rest of the trading period.
The liquidating investor sells throughout the trading period, such that opportunistic investors generate a profit from selling high and buying low.
A larger number of opportunistic investors implies more front-running shortly after~$t=0$.
The liquidating investor reacts accordingly to avoid selling in a falling market and shifts more trading activity to the end.

This pattern holds for many parameter combinations, but it is not universal:
Simulations show that if~$\gamma_1$ is small and~$\rho$ is large, opportunistic investors may start selling again shortly after~$T/2$
such that~$\overline{\alpha}$ changes sign three times.

\subsubsection*{Costs of execution}

\begin{figure}[t]
	\centering
	\begin{tikzpicture}
		\node[anchor=south west,inner sep=0] (image) at (0,0) {\includegraphics[width=0.5\textwidth]{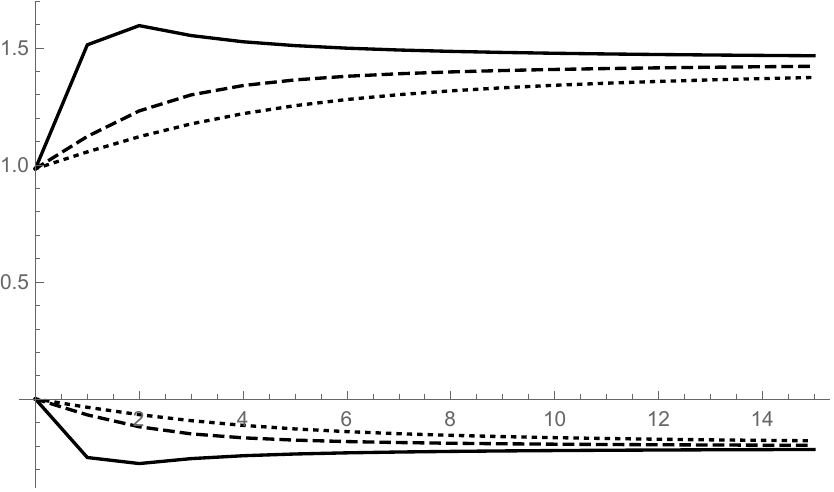}};
	    	\begin{scope}[x={(image.south east)},y={(image.north west)}]
			\node at (0.03,1.07) {Costs};
        		\node at (1.035,0.19) {$n$};
        		\node at (0.8,0.95) {Liquidating investor};
        		\node at (0.8,0.02) {Opportunistic investors (total)};
    		\end{scope}
	\end{tikzpicture}
	\caption{
		Total costs of the liquidating investor and sum of total (negative) costs of the opportunistic investors in dependence of~$n$;
		for~$\gamma_1 = 0.1$ (solid line),~$\gamma_1 = 0.5$ (dashed line) and~$\gamma_1 = 1$ (dotted line).
		Parameters:~$T = 1$, $\rho = 0.1$, $\gamma_0 = 1$, $x_0 = -1$ and $x_1 = 0$.
	}
	\label{oas_fig:npredator_totalcost}
\end{figure}
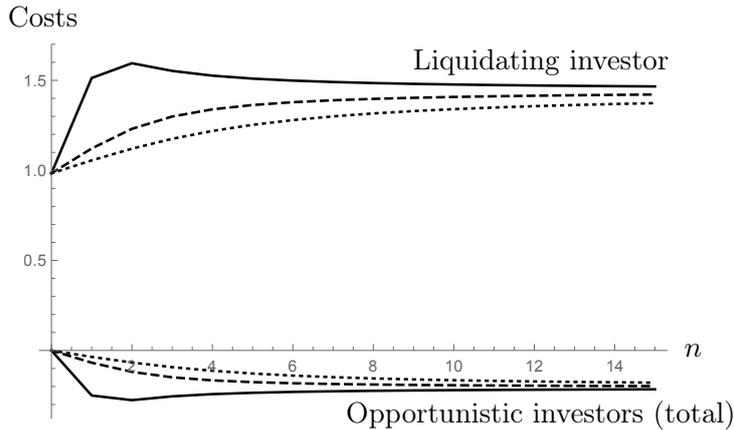

Front-running amplifies the liquidating investor's price impact and increases his total costs significantly (see Figure~\ref{oas_fig:npredator_totalcost}).

If opportunistic investors have low transaction costs, one or two of them suffice to fully realize the profit potential of order anticipation.
Further increasing~$n$ results in competition among opportunistic investors, reducing their total profit 
and also reducing the liquidating investor's total costs.
This is different if transaction costs are high.
Transaction costs restrict the degree to which investors can benefit from opportunistic trading.
High transaction costs leave unrealized profit potential for additional opportunistic investors.
Consequently, the liquidating investor's total costs increase in~$n$.
Notice that while total profits of opportunistic investors generally increase in~$n$, 
these profits are divided (equally) among an increasing number of investors;
each opportunistic investor earns less if~$n$ increases.

It is interesting that the liquidating investor's costs rise significantly if~$\gamma_1$ decreases
(at least for small~$n$), while simulations show that his optimal strategy hardly changes.
This suggests that the liquidating investor can do little to avoid exploitation from order anticipation strategies.

\subsubsection*{Asset price and price overshooting}

\begin{figure}[t]
	\centering
	\begin{tikzpicture}
		\node[anchor=south west,inner sep=0] (image) at (0,0) {\includegraphics[width=0.5\textwidth]{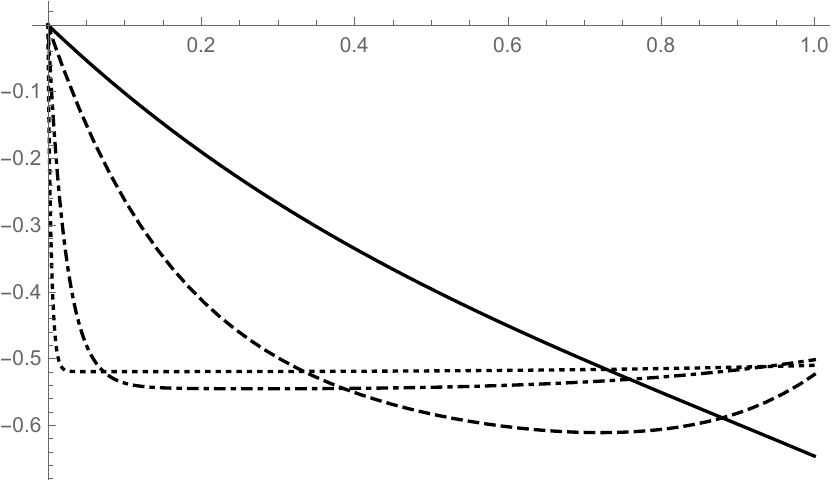}};
	    	\begin{scope}[x={(image.south east)},y={(image.north west)}]
			\node at (0.06,1.05) {$S(t)$};
        		\node at (1.035,0.955) {$t$};
        	\end{scope}
	\end{tikzpicture}
	\caption{
		Asset price~$S(t)$ for~$n=0$ (solid line),~$n=1$ (dashed line),~$n=5$ (dot-dashed line) and~$n=25$ (dotted line).
		Parameters:~$T = 1$, $\rho = 0.95$, $\gamma_0 = 1$, $\gamma_1 = 0.1$, $x_0 = -1$ and $x_1 = 0$.
		}
	\label{oas_fig:npredator_assetprice}
\end{figure}

Figure~\ref{oas_fig:npredator_assetprice} shows that in the absence of opportunistic investors, the asset price decreases steadily over time;
it exhibits a persistent drift.
This changes drastically once opportunistic investors enter the picture,
especially if there are many of them. 
Opportunistic investors build up short positions very quickly, 
causing a sudden price drop right after~$t=0$.
The asset price remains almost constant afterwards.
This seems to support claims about opportunistic investors improving price discovery:
The price drop caused by the order~$x_0$ occurs earlier and more quickly
(see Sections~6.2 and~6.3 in \citealp{Benos2012}, for a discussion).
But inferences about price discovery are outside the scope of this model
because all liquidation constraints are known to all investors.


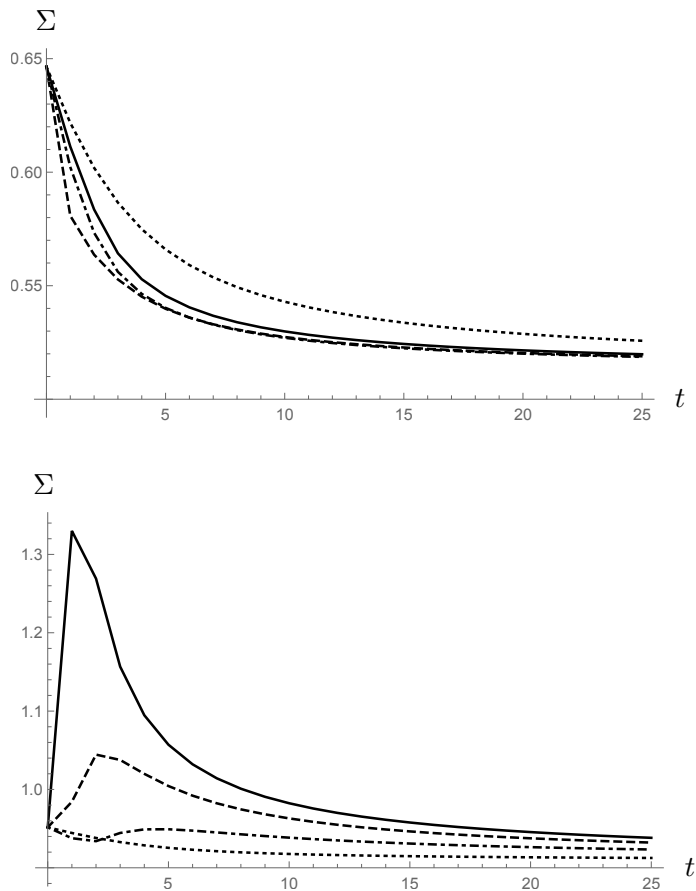
\begin{figure}[t]
	\centering
	\begin{tikzpicture}
		\node[anchor=south west,inner sep=0] (image) at (0,0) {\includegraphics[width=0.5\textwidth]{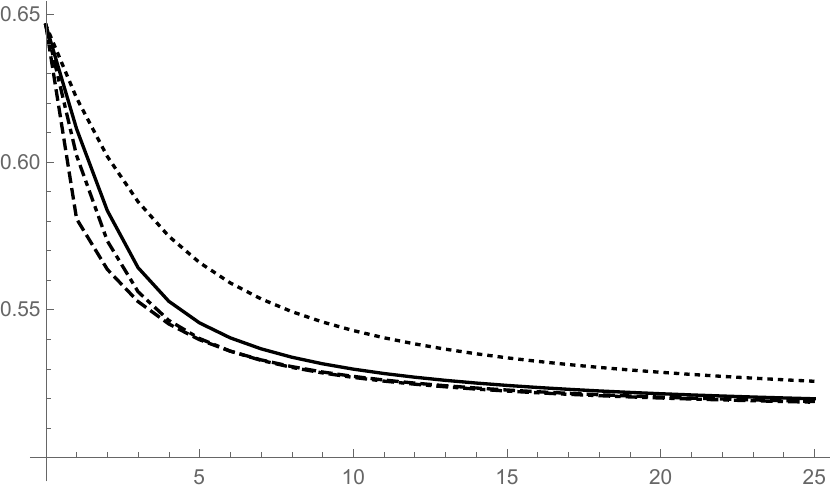}};
	    	\begin{scope}[x={(image.south east)},y={(image.north west)}]
			\node at (0.055,1.07) {$\Sigma$};
        		\node at (1.035,0.075) {$t$};
        	\end{scope}
	\end{tikzpicture}\\
	~\\
	\begin{tikzpicture}
		\node[anchor=south west,inner sep=0] (image) at (0,0) {~ \includegraphics[width=0.5\textwidth]{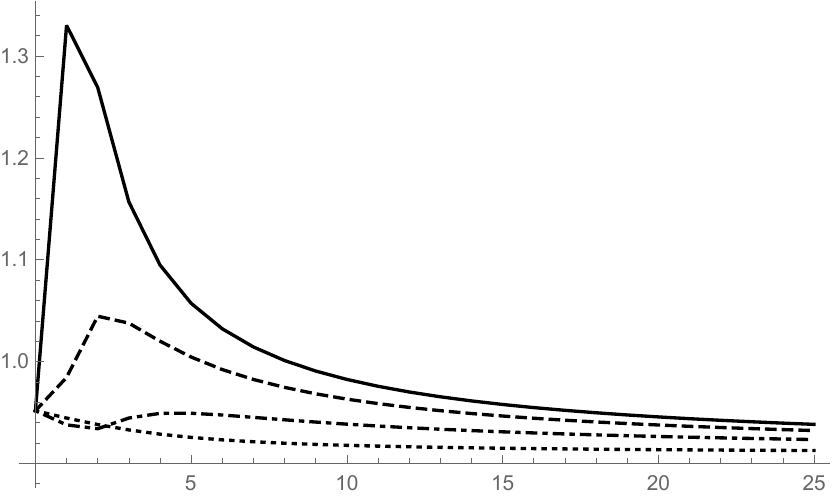}};
	    	\begin{scope}[x={(image.south east)},y={(image.north west)}]
			\node at (0.065,1.07) {$\Sigma$};
        		\node at (1.035,0.072) {$t$};
        	\end{scope}
	\end{tikzpicture}
	\caption{
		Maximum deviation~$\Sigma$ of the asset price for~$\rho = 0.95$ (top) and~$\rho = 0.1$ (bottom)
		and for~$\gamma_1=0.1$ (solid line),~$\gamma_1=0.25$ (dashed line),~$\gamma_1 = 0.5$ (dot-dashed line) and~$\gamma_1=1$ (dotted line).
		Parameters:~$T = 1$, $\gamma_0 = 1$, $x_0 = -1$ and $x_1 = 0.$
		}
	\label{oas_fig:npredator_maxdeviation}
\end{figure}

Consider the maximum deviation of the asset price, 
$$
	\Sigma \coloneqq \sup_{t \in [0,T]} |S(t) - S(0)|.
$$
\cite{Brunnermeier2005} claim that opportunistic investors cause price overshooting, 
i.e., that~$\Sigma$ is larger for~$n \ge 1$ than for~$n=0$.
They argue that this may lead to a domino effect:
The price drop caused by the liquidating investor and amplified by the opportunistic investors triggers additional sell orders
(for instance from pending stop-loss orders).
This causes an even more extreme price drop, triggering further sell orders, etc.

The model of price impact in which Brunnermeier and Pedersen observe price overshooting 
only features temporary and permanent price impact.
Both impose few constraints on the opportunistic traders
and generate little feedback from opportunistic traders to the liquidating trader.
Consequently, opportunistic traders trade aggressively and scale their strategies to an exogenously given maximum size.
Brunnermeier and Pedersen arrive at a grim picture in which ``predators'' (opportunistic traders) exploit
``distressed traders'' (liquidating investors) and may even cause a ``panic'' (the domino effect described above).
Price overshooting is also observed by \cite{Oehmke2014}, again in a model with temporary and permanent price impact only.

Figure~\ref{oas_fig:npredator_maxdeviation} shows that in the present model, price overshooting is the exception, not the rule:
In general,~$\Sigma$ decreases if~$n$ increases.
This is most evident for markets with a ``short memory'', i.e., for large values of~$\rho$.
A possible explanation is that the price overshooting observed by Brunnermeier and Pedersen 
is a consequence of permanent (or long-lived transient) price impact,
rather than an inherent consequence of opportunistic trading.

Another possible explanation is that quadratic transaction costs prevent price overshooting.
Quadratic costs imply that a (statistical) arbitrage strategy cannot be scaled indefinitely without becoming unprofitable.
As~$n$ increases, competition among opportunistic investors increases, and transaction costs increasingly work against them.
Figure~\ref{oas_fig:npredator_maxdeviation} shows, however, that there is no obvious relationship between~$\Sigma$ 
and the level~$\gamma_1$ of transaction costs.

\section{Heterogeneous time horizons}
\label{oas_section:extensions}

Opportunistic investors do not necessarily have the same time horizon as the liquidating investor.
\cite{Admati1991} point out that the liquidating investor may preannounce his liquidation constraint:
``By informing potential traders who can take the other side of the preannounced orders and by
allowing the market to prepare to absorb these orders, preannouncement facilitates the match between
the demand and supply of liquidity in the market'' (p.~444).
This practice is known as sunshine trading.
It can be implemented by demanding that the liquidating investor only trade after some time~$T_0 > 0,$
as in \cite{Brunnermeier2005}.

In their analysis of sunshine trading, \cite{Schoneborn2009} argue that opportunistic investors may also have additional time to unwind their position 
after the liquidating investor has fully executed his order.
This can be implemented in the model by demanding that the liquidating investor only trade until some time~$T_1 < T$.

Consequently, divide~$[0,T]$ into three periods:
The \textit{acquisition period}~$[0,T_0]$, the \textit{main period}~$[T_0, T_1]$ and the \textit{unwinding period}~$[T_1, T].$
Suppose for now that~$T_0$ and~$T_1$ are fixed.
The liquidating investor is only allowed to trade during the main period.
Opportunistic investors begin and end with a flat inventory~$X_1(0) = X_1(T) = 0.$ 
They use the acquisition period to build up a position~$X_1(T_0)$, then trade alongside the liquidating investor during the main period. 
At the end of the main period, they hold a position~$X_1(T_1)$, which they liquidate in the unwinding period.
Notice that in equilibrium, all opportunistic investors still behave identically.

Given~$X_1(T_0),$ the acquisition period is described by the model in Section~\ref{oas_chapter:existence}, 
where investors~$i=1, 2, \dotsc, n$ acquire identical amounts~$x_1 = x_2 = \dotsb = x_n = X_1(T_0)$ over the time horizon~$[0,T_0]$.
Theorem~\ref{oas_th:ode} yields the equilibrium strategies.

During the main and the liquidation period, the situation is more complicated.
The model by Sch{\"o}neborn and Schied features linear temporary and permanent price impact.
This has the computational advantage that price impact generated during earlier periods does not affect equilibrium strategies in subsequent periods.
With transient price impact, trades from earlier periods cause a (deterministic) price drift in subsequent periods.
During the main period, the asset price becomes
$$
	S(t) = e^{-\rho (t-T_0)} S(T_0) + \int_{T_0}^t e^{-\rho(t-s)} \sum_{i=0}^n \alpha_i(s) \dif s, \qquad t \in [T_0, T_1].
$$
In the same way, price impact from the main period generates a price drift during the liquidation period.
Theorem~\ref{oas_th:ode} must be generalized by replacing~$S$ with
$$
	\tilde S(t) \coloneqq e^{-\rho (t-\tau_0)} s + S(t), \qquad t \in [\tau_0, \tau_1],
$$
for~$(\tau_0, \tau_1) \in \{(T_0,T_1), (T_1,T)\}$ and~$s \in \mathbb{R}.$
Repeating the arguments from Section~\ref{oas_chapter:existence}, one sees 
that the Nash equilibrium~$\psi^*$ 
still satisfies a system of differential equations of the form~$\frac{\dif}{\dif t} \psi^* = M \psi^* + m$, but now~$m = m(t)$ is not constant.
Once this system is solved, one may calculate the optimal strategies for the liquidating investor during the main period, 
and for the opportunistic investors during the main and liquidation periods, in dependence of~$X_1(T_0)$ and~$X_1(T_1).$
This yields the total (negative) costs for opportunistic investors over the entire time horizon~$[0,T]$ 
in dependence of~$X_1(T_0)$ and~$X_1(T_1)$.
It remains to minimize these costs over~$(X_1(T_0), X_1(T_1)) \in \mathbb{R}^2$.

But go one step further.
In the current setting, a liquidating investor engaging in sunshine trading may not only announce his liquidation constraint~$x_0$, 
but also his time horizon~$[T_0, T_1].$
Sch{\"o}neborn and Schied show that a shorter trading horizon~$T_1 < T$ can be beneficial to the liquidating investor in certain market conditions.
Although there may be an exogenous upper bound on~$T_1$, it is reasonable to assume 
that the liquidating investor can voluntarily commit to a shorter trading horizon. 
The liquidating investor also has some control over~$T_0$ because he can choose the time of announcement~$t=0$.
Hence~$T_0$ and~$T_1$ should not be seen as fixed. 
One should rather perform a final optimization over~$(T_0,T_1) \in \mathbb{R}^2$, 
this time minimizing the liquidating investor's total costs during the main period.

This extension promises interesting results, with opportunistic investors possibly engaging in liquidity provision 
instead of front-running, as observed by~\cite{Schoneborn2009}.

\section{Conclusion}

Order anticipation strategies try to benefit from a large order's price impact.
Since these strategies initially trade in the same direction as the large order, 
\cite{Brunnermeier2005} suspect them to cause price overshooting.
Linking opportunistic trading with increased (or reduced) price overshooting 
would serve as a strong argument for (or against) the harmfulness of order anticipation strategies.

This paper analyzes order anticipation strategies in a theoretical framework.
A multi-investor model of optimal execution under transient price impact is introduced and shown to admit a unique Nash equilibrium 
if trading incurs quadratic transaction costs.
For the special case of exponentially decaying price impact, the Nash equilibrium is derived in closed form.
Numerical simulations show that while order anticipation strategies raise the costs of executing a large order significantly,
price overshooting typically does not occur.

The current model differs from the model by \cite{Brunnermeier2005} in two ways: Price impact is transient, not permanent; 
and trading incurs quadratic transaction costs.
It is an open question which of these differences is responsible for the absence of price overshooting.

Further theoretical research should determine for what types of price impact and transaction costs price overshooting occurs;
and empirical research is necessary to study the influence of order anticipation strategies in real financial markets.

\appendix
\section{Proofs}

\begin{prf}[Proof of Lemma~\ref{oas_lemma:fredholm}.]~\\
	Define the linear subspace~$\mathcal{B} \coloneqq \{ \beta \in L^2[0,T] \mid \int_0^T \beta(t) \dif t = 0 \}.$

	Necessity:
	Suppose~$\alpha^* = (\alpha^*_0, \alpha^*_1, \dotsc, \alpha^*_n)$ is a Nash equilibrium
	in the class of deterministic strategies.
	Let~$i = 0, 1, \dotsc, n.$
	For every~$y \in \mathbb{R}$ and~$\beta \in \mathcal{B},$
	the function~$\alpha_i^* + y \beta$ is a deterministic admissible strategy for investor~$i$.
	It follows that a necessary condition for the optimality of~$\alpha^*_i$ is 
	\begin{align*}
		0 &= \left. \frac{\dif}{\dif y} J_i [ \alpha^*_i + y \beta \,|\, \alpha^*_{-i} ] \right|_{y = 0}\\
		&= \int_0^T \Big( \beta(t) \big( \gamma_i \alpha^*_i(t) +S(t; \alpha^*) \big) + \int_0^t G(t-s) \alpha^*_i(t) \beta(s) \dif s \Big) \dif t.
	\end{align*}
	Conclude with Fubini's theorem that
	\begin{align*}
		0 &= \int_0^T \beta(t) \Big( \gamma_i \alpha^*_i(t) + \int_0^t G(t-s) \sum_{i=0}^n \alpha^*_i(s) \dif s 
			+ \int_t^T G(s-t) \alpha^*_i(s) \Big) \dif s \\
		&= \int_0^T \beta(t) (F \alpha^*)_i (t) \dif t.
	\end{align*}
	The fundamental lemma of the calculus of variations (see, e.g., Lemma~2 in \citealp{Gelfand1963}) 
	implies that~$(F \alpha^*)_i$ is constant for almost all~$t \in [0,T]$.

	Sufficiency: 
	Let~$\alpha^* = (\alpha^*_0, \alpha^*_1, \dotsc, \alpha^*_n)$ be such that for every~$i = 0, 1, \dotsc, n,$
	the liquidation constraint~$\int_0^T \alpha^*_i(t) \dif t = x_i$ is satisfied and~$(F \alpha^*)_i (t) = \eta_i$ for almost all~$t \in [0, T]$
	for some~$\eta = (\eta_0, \eta_1, \dotsc, \eta_n) \in \mathbb{R}^{n+1}.$
	
	Let~$i = 0, 1, \dotsc, n.$
	The liquidation constraint implies that any deterministic admissible strategy~$\alpha_i$ for investor~$i$ 
	can be written as~$\alpha_i = \alpha^*_i + \beta$ 
	for some~$\beta \in \mathcal{B}$.
	Conclude with Fubini's theorem that
	\begin{align*}
		J_i [ \alpha_i \,|\, \alpha^*_{-i} ]
		&= J[\alpha^*_i \,|\, \alpha^*_{-i} ] + \eta_i \int_0^T \beta(t) \dif t \\
		&\hphantom{{}=}{}+ \frac12 \int_0^T \Big( \gamma_i \beta(t)^2 + \int_0^T G(|t-s|) \beta(t) \beta(s) \dif s \Big) \dif t  \\
		&\ge J[\alpha^*_i \,|\, \alpha^*_{-i}].
	\end{align*}
	Hence~$\alpha^*_i$ is the optimal strategy for investor~$i$ given that the other investors pursue~$\alpha^*_{-i}.$
	This is true for all~$i$, showing that~$\alpha^*$ is a Nash equilibrium.

	Furthermore, for all~$i = 0, 1, \dots, n,$
	\begin{align*}
		J_i[ \alpha^*_i \,|\, \alpha^*_{-i}] 
		&= \int_0^T \Big( \frac{\gamma_i}{2} \alpha^*_i(t)^2 
			+ \alpha^*_i(t) \int_0^t G(t-s) \sum_{j =0}^{n} \alpha^*_j(s) \dif s \Big) \dif t \\
		&= \int_0^T \alpha^*_i(t) (F \alpha^*)_i(t) \dif t \\
		&\hphantom{{}=}{}- \frac12 \int_0^T \alpha^*_i(t) \Big( \gamma_i \alpha^*_i(t) + \int_0^T G(|t-s|) \alpha^*_i(s) \dif s \Big) \dif t \\
		& \le \eta_i x_i.
	\end{align*}
\qed\end{prf}

\begin{prf}[Proof of Theorem~\ref{oas_th:existence}.]~\\
	Uniqueness is established by Lemma~\ref{oas_lemma:zhang}(i).

	Define~$F$ as in~\eqref{oas_eq:F}.
	Let~$\langle\cdot,\cdot\rangle$ denote the $L^2$-inner product on $[0,T]$,
	and~$\lVert\cdot\rVert$ its induced norm.
	Without loss of generality, assume~$\gamma_0 \le \gamma_i \le \gamma_n$ for all~$i = 0, 1, \dotsc, n.$
	
	Conclude with the Cauchy-Schwarz inequality and Jensen's inequality that~$F$ is bounded:
	For every~$\alpha \in L^2([0,T]; \mathbb{R}^{n+1}),$
	\begin{align*}
		&\hphantom{\le} \| F\alpha \| \\
		&\le \| \Gamma \alpha \| + \Big( \int_0^T \Big(\int_0^t G(t-s) \sum_{i=0}^n \alpha_i(s) \dif s \Big)^2 (n+1) \dif t \Big)^{1/2}\\
		&\hphantom{{}\le}{}+ \Big( \sum_{i=0}^n \int_0^T \Big( \int_t^T G(s-t) \alpha_i(s) \dif s \Big)^2 \dif t \Big)^{1/2} \\
		&\le \gamma_n \| \alpha \| + (n+2) \Big( \int_0^T \int_0^T G(|t-s|)^2 \dif s \dif t \Big)^{1/2} \| \alpha \|.
	\end{align*}
	Recall that~$G$ is square-integrable. 
	Furthermore, apply Fubini's theorem and recall that~$G$ is of positive type to see that
	\begin{align*}
		\langle F\alpha, \alpha \rangle &=  
			\sum_{i=0}^n \gamma_i \int_0^T \alpha_i(t)^2 \dif t \\
		&\hphantom{{}=}{}+ \frac12 \int_0^T \int_0^T G(|t-s|) \sum_{i=0}^n \alpha_i(t) \sum_{i=0}^n \alpha_i(s) \dif s \dif t \\
		&\hphantom{{}=}{}+ \frac12 \sum_{i=0}^n \int_0^T \int_0^T G(|t-s|) \alpha_i(t) \alpha_i(s) \dif s \dif t \\
		& \ge \gamma_0 \| \alpha \|^2.
	\end{align*}
	Hence~$F$ is also bounded from below.
	The adjoint~$F^*$ of~$F$ is given by
	$$
		(F^*\alpha)(t) = \Gamma \alpha(t) + \Big(\int_t^T G(s-t) \alpha(s)^\top \one(s) \dif s\Big)\, \one(t) + \int_0^t G(t-s) \alpha(s) \dif s.
	$$
	The same arguments as before show that~$F^*$ is bounded from above and below.
	Conclude that~$F$ is invertible:
	Boundedness from below implies $\text{ker}(F) = \{{\bm 0}\}$, showing that $F$ is injective;
	and $\text{range}(F) = \text{ker}(F^*)^\perp = \{{\bm 0}\}^\perp,$ showing that $F$ is surjective.
	
	Now define a linear operator~$A \colon \mathbb{R}^{n+1} \to \mathbb{R}^{n+1}$ via
	$$
		A \eta \coloneqq \int_0^T (F^{-1}{\bm \eta})(t) \dif t.
	$$
	
	It follows from the calculations above that 
	$$
		\eta^\top A \eta = \langle {\bm\eta}, F^{-1}{\bm\eta}\rangle \ge \gamma_0 \lVert F^{-1}{\bm \eta}\rVert^2.
	$$
	Since $\text{ker}(F^{-1}) = \{{\bm 0}\},$ conclude that $\eta^\top A \eta = 0$ if and only if $\eta = (0, 0, \dotsc, 0).$
	Hence $A$ is invertible.
	
	For given liquidation constraints~$x,$
	define~$\eta \coloneqq A^{-1} x$ and~$\alpha^* \coloneqq F^{-1} {\bm \eta}.$
	Then~$\alpha^*$ is a Nash equilibrium in the class of deterministic strategies by Lemma~\ref{oas_lemma:fredholm}: 
	First,~$\int_0^T \alpha^*(t) \dif t = A \eta = x$;
	second,~$F \alpha^* = {\bm \eta}$.
	By Lemma~\ref{oas_lemma:zhang}, it is the unique Nash equilibrium in the class of admissible strategies.
\qed\end{prf}

\begin{prf}[Proof of Theorem~\ref{oas_th:ode}.]~\\
	A function~$\psi = (\alpha_0, \alpha_1, \dotsc, \alpha_n, S) \in L^2([0,T]; \mathbb{R}^{n+2})$ shall be called \textit{regular} 
	if for each~$i = 0, 1, \dotsc, n,$ the function~$\alpha_i$ is a deterministic admissible strategy for investor~$i,$ 
	and~$S = S(\cdot\,;\alpha)$ is the asset price corresponding to~$\alpha \coloneqq (\alpha_0, \alpha_1, \dotsc, \alpha_n).$

	Define the~$(n+1)\times(n+2)$-dimensional matrices
	$$
		U \coloneqq \begin{pmatrix} 
			\gamma_0 & 0 & \dots & 0 & 1 \\
			0 & \gamma_1 & \dots & 0 & 1 \\
			\vdots & \vdots & \ddots & \vdots & \vdots \\
			0 & 0 & \dots & \gamma_n & 1
			\end{pmatrix}
		\quad \text{ and } \quad\,
		V \coloneqq \begin{pmatrix}
			\frac{\rho}{\gamma_0} & 0 & \dots & 0 & 0 \\
			0 & \frac{\rho}{\gamma_1} & \dots & 0 & 0 \\
			\vdots & \vdots & \ddots & \vdots & \vdots \\
			0 & 0 & \dots & \frac{\rho}{\gamma_n} & 0
		\end{pmatrix}.
	$$
	
	The proof is in four steps.
	
	1. \textit{$\psi$ is a Nash equilibrium if and only if it is regular, continuously differentiable and solves the system of differential equations
		\begin{equation}
		\label{oas_eq:matrixode}
			\frac{\dif}{\dif t} \psi = M \psi - V^\top y,
		\end{equation}
		where~$y = U \psi(T).$}
	
		Necessity: 
		Let~$\psi$ be a Nash equilibrium.
		It is clear from~\eqref{oas_eq:fredholmexp} and~\eqref{oas_eq:Sderivative} that~$\psi$ is continuously differentiable.
		Let~$t=T$ in~\eqref{oas_eq:fredholmexp} to obtain~$y = U \psi(T).$
		From~\eqref{oas_eq:secondorderode} and~\eqref{oas_eq:Sderivative}, deduce~\eqref{oas_eq:matrixode}.
		
		Sufficiency:
		Let~$\psi = (\alpha_0, \alpha_1, \dotsc, \alpha_n, S)$ be regular and suppose it solves~\eqref{oas_eq:matrixode} with~$y = U \psi(T).$
		Let~$\alpha \coloneqq (\alpha_0, \alpha_1, \dotsc, \alpha_n).$
		Fix~$i = 0, 1, \dotsc, n.$
		Define
		$$
			f_i(t) \coloneqq \int_t^T e^{\rho(t-s)} \alpha_i(s) \dif s.
		$$
		Clearly, the derivative~$f_i'$ satisfies~$f_i' = \rho f_i - \alpha_i.$
		Let~$
			g_i \coloneqq \eta_i - \gamma_i \alpha_i - S.
		$
		With~\eqref{oas_eq:matrixode} conclude~$g_i' = \rho g_i - \alpha_i$.
		Hence~$f_i' - g_i' = \rho (f_i - g_i).$
		Applying the boundary condition~$y = U \psi(T)$ shows that~$f_i(T) - g_i(T) = 0$.
		It follows that~$f_i = g_i$ and
		$$
			(F\alpha)_i = \gamma_i \alpha_i + S + f_i = \gamma_i \alpha_i + S + g_i = \eta_i.
		$$
		This is true for all~$i = 0, 1, \dotsc, n.$
		Hence~$\psi$ is a Nash equilibrium by Lemma~\ref{oas_lemma:fredholm}.
		
	2. \textit{$M$ is invertible.}
	
		Define the~$(n+2)$-dimensional column vectors
		\begin{align*}
			v_1 &\coloneqq \Big(\rho + \frac1{\gamma_0}, \rho + \frac1{\gamma_1}, \dotsc, \rho + \frac1{\gamma_n}, -\frac12\Big),
				&v_2 &\coloneqq \Big(\hspace{-.15cm} -\frac1{\gamma_0}, -\frac1{\gamma_1}, \dotsc, -\frac1{\gamma_n}, 1\Big), \\
			v_3 &\coloneqq (1, 1, \dotsc, 1, -2\rho),
				&u &\coloneqq (1, 1, \dotsc, 1).
		\end{align*}
		Then~$M = (\diag(v_1) + v_2 u^\top ) \diag(v_3)$ and, by the matrix determinant lemma,
		$$
			\det M = - \rho \Big( 1 + \sum_{i=0}^n \frac1{\rho \gamma_i + 1}\Big) \prod_{i=0}^n \Big( \rho + \frac1{\gamma_i}\Big) \neq 0.
		$$
	
	3. \textit{$\psi$ is a Nash equilibrium if and only if it is regular and there is a~$z \in \mathbb{R}^{n+2}$ such that}
		\begin{equation}
		\label{oas_eq:NEform}
			\psi(t) = (e^{Mt} + N_1 e^{MT}) z.
		\end{equation}
		The general solution of~\eqref{oas_eq:matrixode} is~$e^{Mt} z + M^{-1} V^\top y$ for~$z \in \mathbb{R}^{n+2}.$
		By Step~1, the condition~$y = U (e^{MT} z + M^{-1} V^\top  y)$ must be satisfied.
		The matrix~$M - V^\top U$ is nonsingular, which can be verified by checking that~$(M - V^\top U)^{-1} = (N_1 - v v^\top) / \rho.$
		By the Woodbury matrix identity,
		$$
			(\Id_{n+1} - U M^{-1} V^\top )^{-1} = \Id_{n+1} + U(M-V^\top  U)^{-1} V^\top .
		$$
		Hence 
		$$
			y = ( \Id_{n+1} + U(M - V^\top U)^{-1} V^\top ) U e^{MT} z.
		$$
		It holds that~$V^\top  U(\Id_{n+2} + N_1) = M N_1$, or equivalently,~$(M - V^\top U)^{-1} V^\top U = N_1.$
		With this, obtain
		$$
			M^{-1} V^\top  y = M^{-1} V^\top U (\Id_{n+2} + N_1) e^{MT} z = N_1 e^{MT} z.
		$$

	4. Once~$\psi$ is defined by~\eqref{oas_eq:NEform}, integrating over~$[0,T]$ shows that~$\psi$ is regular if and only if~$N_2 z = \tilde{x}.$
		It remains to show that~$N_2$ is invertible.
		Although this follows from Theorem~\ref{oas_th:existence}, a separate proof is given here.
		
		Consider the case where all investors must trade zero net amounts, i.e.,~$x = (0, 0, \dotsc, 0)$.
		It is easy to check that in this case,~$\psi^0 \coloneqq \zero$ is a Nash equilibrium.
		By Lemma~\ref{oas_lemma:zhang}(ii), this is the only Nash equilibrium.
		According to Step~3, there exists a~$z^0 \in \mathbb{R}^{n+2}$ such that~$\psi^0(t) = (e^{Mt} + N_1 e^{MT})z^0$.
		This shows that~$(0, 0, \dotsc, 0) = \frac{\dif}{\dif t} \psi^0(t) = M e^{Mt} z^0$ for all~$t \in [0,T].$
		The matrix~$M e^{Mt}$ is invertible, hence~$z^0 = (0, 0, \dotsc, 0).$
		It follows that the equation~$N_2 z = (0, 0, \dotsc, 0)$ has only the trivial solution~$z^0 = (0, 0, \dotsc, 0)$, so~$N_2$ is invertible.
\qed\end{prf}

\bibliography{}

\end{document}